\documentclass{INTERSPEECH2023}

\interspeechcameraready

\title{PromptStyle: Controllable Style Transfer for Text-to-Speech with Natural Language Descriptions}
\name{Guanghou Liu$^1$, Yongmao Zhang$^1$, Yi Lei$^1$, Yunlin Chen$^2$, Rui Wang$^2$,  Zhifei Li$^2$, Lei Xie$^{1*}$\thanks{* Corresponding author.}}
\address{
  $^1$Audio, Speech and Language Processing Group (ASLP@NPU)\\School of Computer Science,
  Northwestern Polytechnical University, Xi’an, China\\
  $^2$Shanghai Mobvoi Information Technology Co., Ltd}
\email{ghliu@mail.nwpu.edu.cn, yunlinchen@mobvoi.com, lxie@nwpu.edu.cn}

\usepackage{subfigure}
\usepackage{multirow}
\usepackage{multicol}
\usepackage{array}

\begin{document}

\maketitle
 
\begin{abstract}
Style transfer TTS has shown impressive performance in recent years. However, style control is often restricted to systems built on expressive speech recordings with discrete style categories. In practical situations, users may be interested in transferring style by typing text descriptions of desired styles, without the reference speech in the target style. The text-guided content generation techniques have drawn wide attention recently. In this work, we explore the possibility of controllable style transfer with natural language descriptions. To this end, we propose \textit{PromptStyle}, a text prompt-guided cross-speaker style transfer system. Specifically, PromptStyle consists of an improved VITS and a cross-modal style encoder. The cross-modal style encoder constructs a shared space of stylistic and semantic representation through a two-stage training process. Experiments show that PromptStyle can achieve proper style transfer with text prompts while maintaining relatively high stability and speaker similarity.\renewcommand{\thefootnote}{\arabic{footnote}} Audio samples are available in our demo page\footnote[1]{ \url{https://PromptStyle.github.io/PromptStyle}}.

\end{abstract}
\noindent\textbf{Index Terms}: text-to-speech, style transfer, style prompt

\vspace{2pt}
\section{Introduction}
\label{sec:intro}

Text-to-speech (TTS)~\cite{Wang2017Tacotron,ren2020fastspeech} aims to produce human-like speech from input text. Recent progress in deep learning approaches has greatly improved the naturalness of speech~\cite{tan2021survey}. With the wide applications of TTS in real-world human-computer interaction, expressive TTS with diverse styles attracts more attention. Generating stylistic speech for a specific speaker intuitively needs the same speaker's high-quality expressive speech recordings, which incurs a high cost for data collection. To solve the problem of synthesizing expressive speech for the target speaker without diverse speaking styles, cross-speaker style transfer~\cite{Bian2019Multi,Li2021ControllableCE,zhang2022iemotts,li22h_interspeech,shang2021incorporating} is a feasible solution.

%However, in practice, high-quality expressive data usually contains only a small number of speaker due to the cost of data recording, which greatly limits the application of speech synthesis technology. In order to generate highly expressive speech for any speaker, it is necessary to transfer a given speech style to the target speaker that exclude the target style.

For the style representations in style transfer scenarios, existing works mainly include two different methods, i.e. the pre-defined style id category index~\cite{Li2018EMPHASISAE,Zhu2019ControllingES,tits2019visualization,tits2020exploring} and hidden variables extracted from the reference signal~\cite{Skerry2018Towards,Zhang2019LearningLR,Kulkarni2020IMPROVINGLR,Jia2018TransferLF,Li2021ControllableET}. However, the id-based methods are limited to the styles of fixed discrete categories, which leads to less flexibility. Although the reference-based methods can produce various speech through different references, the extracted style representation is not interpretable. Moreover, in practical applications, it is difficult to accurately and conveniently select an appropriate reference for arbitrary textual content.
%We observe that there are generally two types of methods to control the style of speech in a style transfer task. One uses a few pre-defined style classification index as condition of speech synthesis, and the other extracts style information from reference audio to guide the synthesis. However, the diversity of the former is limited as the model can just generate preset styles. Although the latter generates variety of expressive speech through different references, the extracted style information from references is not intuitive and interpretable, and a reference speech should be provided in each infer process. In practical applications, it is difficult for users to accurately and conveniently select appropriate audio for reference. In practice, It is difficult for users to select the reference voice they want accurately and conveniently.

With the success of text and image generation from prompt descriptions~\cite{zhang2021cpm,seo2022end}, some prompt-based TTS methods are proposed to improve the expressiveness and naturalness of synthetic speech. Style-Tagging-TTS (ST-TTS)~\cite{kim2021expressive} produces expressive speech based on style tags, which are stylistic words or phrases labeled from audiobook datasets. But it is difficult to describe complex styles by a single word or phrase, and the style tags are hard to label in common audiobook datasets as most utterances may be recorded in a less-expressive reading style. PromptTTS~\cite{guo2022prompttts} proposes to use a style prompt from five different factors (i.e. gender, pitch, speaking speed, volume, and emotion) to guide the style expression for the generated speech. The recent InstructTTS~\cite{yang2023instructtts} can synthesize stylistic speech with the guidance of natural language descriptions without formal constraints as style prompts. It's a three-stage training approach to capture semantic information from natural language style prompts as conditioning to the TTS system. Intuitively, natural language description is a convenient and user-friendly way to describe the desired style since no prior acoustic knowledge is required. 

In this study, we focus on style transfer in the audiobook generation, where the target speaker has little expressive data and no style description prompts. Through other expressive data and transferring diversified styles with natural language descriptions, expressive audiobook speech can be generated for the target speaker. Specifically, this paper proposes to leverage natural language description prompts to transfer style from the source speaker to the target speaker who has no expressive speech data. The proposed approach - \textit{PromptStyle} - has a two-stage procedure utilizing text prompts to model the style appearance. In the first stage, based on an improved VITS~\cite{kim2021conditional}, we use a style encoder to extract a style hidden representation from reference speech as the condition of the TTS system. Multi-speaker multi-style expressive data without style annotation is involved in this stage to achieve cross-speaker style transfer through diverse reference speech. In the second stage, we design a prompt encoder to model the style embedding from the style prompt. Expressive speech data with style annotations in natural language descriptions is involved to fine-tune the pre-trained language model and TTS acoustic model, capturing the relationship between prompt embedding and style embedding space. Due to the generalization capability of the language model, style transfer from unseen prompts is also feasible.

\begin{figure*}[ht]
\centering
\begin{minipage}[b]{0.55\linewidth}
\centering
    \subfigure[Training process]{\includegraphics[scale=0.55]{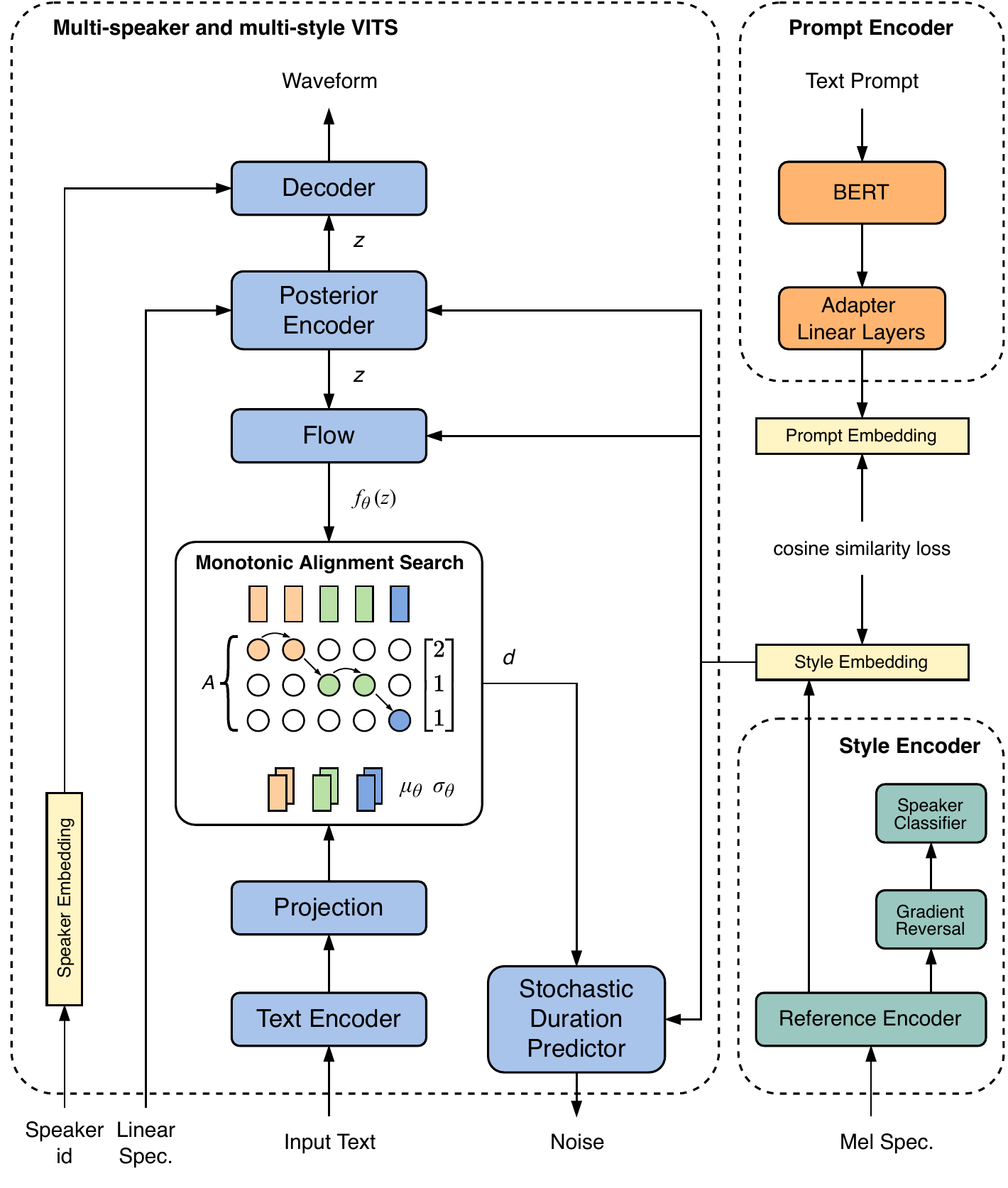}}
\end{minipage}
\centering
\begin{minipage}[b]{0.4\linewidth}
    \subfigure[Inference process]{\includegraphics[scale=0.55]{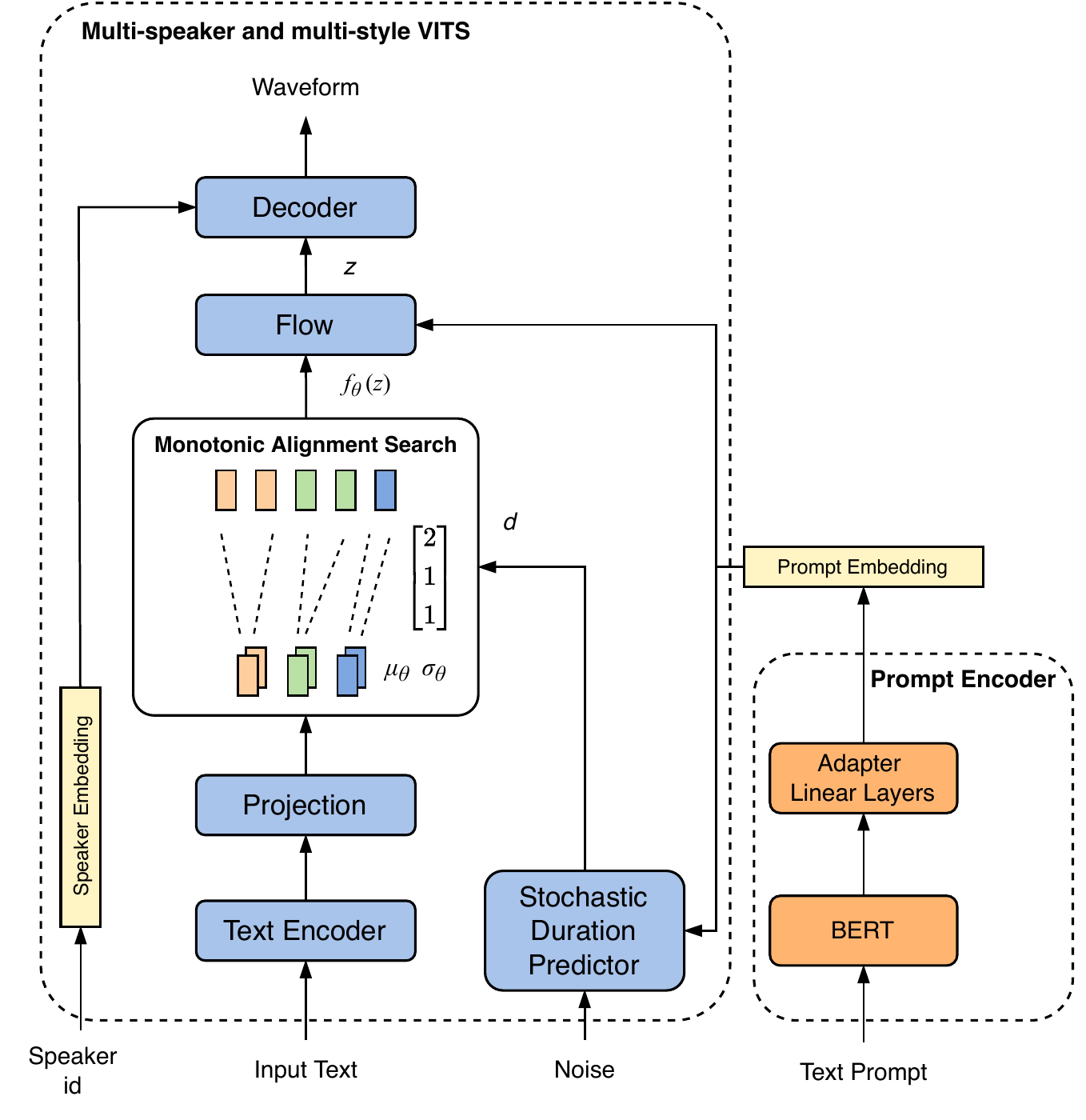}}
\end{minipage}\\
\caption{Architecture of PromptStyle}
 \label{fig:architecture overview}
 \vspace{-0.4cm}
\end{figure*}

\vspace{0.015cm}
We summarize the contributions of PromptStyle as follows.
\begin{itemize}
\item We propose a two-stage TTS approach for cross-speaker style transfer with natural language descriptions, which is more user-friendly and controllable than previous works.
\item The proposed two-stage approach first uses a large amount of data without annotations to train a reference-based style transfer TTS model, and then leverages only a small amount of labeled data with style prompts to fine-tune a prompt encoder and the acoustic model. 
\item With the generalization capability of the pre-trained language model, our approach can generate stylistic speech for the target speaker from an unseen style prompt.
\end{itemize}

Experiments show that our proposed method can conduct cross-speaker style transfer with flexible text prompts and achieve high speech quality.

\vspace{-0.2cm}
\section{PromptStyle}
\label{sec:methods}
\vspace{-0.1cm}

\subsection{Model Overview}
\vspace{-0.1cm}

The overall architecture of the model we proposed is shown in Fig~\ref{fig:architecture overview}. The proposed model consists of three parts, including a style encoder, a prompt encoder, and a TTS system. Our TTS framework is based on VITS[x], an end-to-end TTS model that directly converts the phoneme sequence to the speech waveform. VITS can generate high-quality speech attributed to its end-to-end learning manner without the possible mismatch between the acoustic model and the vocoder in the conventional paradigm. The style encoder is used to extract style information from reference speech and the prompt encoder is used to extract semantic information from natural language descriptions. 

Based on the VITS framework, we propose a two-stage training procedure to achieve style transfer with text prompts. In the first stage, multi-style expressive data with speaker annotations are used to train a style transfer system based on VITS with the style encoder, aiming to learn a style embedding space. In the second stage, we use expressive speech with natural language description prompts to fine-tune the whole model to construct the relationship between the semantic embedding space and the style embedding space, so that we can achieve style transfer with text prompts.

\vspace{-0.1cm}
\subsection{Cross-modal Style Encoder}
\vspace{-0.1cm}

We introduce a cross-modal style encoder consisting of a style encoder and a prompt encoder for constructing a shared space of stylistic and semantic representation. 

The style encoder is used to extract the style information from Mel-spectrograms but not linear spectrograms, because linear spectrograms contain too much information which may prevent capturing the style information accurately. The style encoder consists of a reference encoder and an auxiliary adversarial component. The reference encoder almost has a similar structure to the Global Style Tokens (GSTs)~\cite{Wang2018Style} discarding style token. The auxiliary adversarial component consists of a gradient reversal layer, feed-forward layers, and a speaker classifier, to disentangle the speaker and style information. The style encoder aims to construct a style embedding space for style transfer from large amounts of expressive speech without style annotations.

A prompt encoder is used to build a prompt embedding space for controllable style transfer with natural language descriptions. The prompt encoder consists of a pre-trained BERT~\cite{li2019neural} and an adapter layer. First, the text prompt is fed into the BERT model to extract semantic features. Then the adapter layer converts the semantic features into the prompt embedding of the same size as style embedding from the style encoder, and captures the relationship between the prompt embedding and the style embedding space.

\begin{figure*}[t]
\centering
\begin{minipage}{0.27\linewidth}
\centering
    \subfigure[CST-TTS]{\includegraphics[scale=0.35]{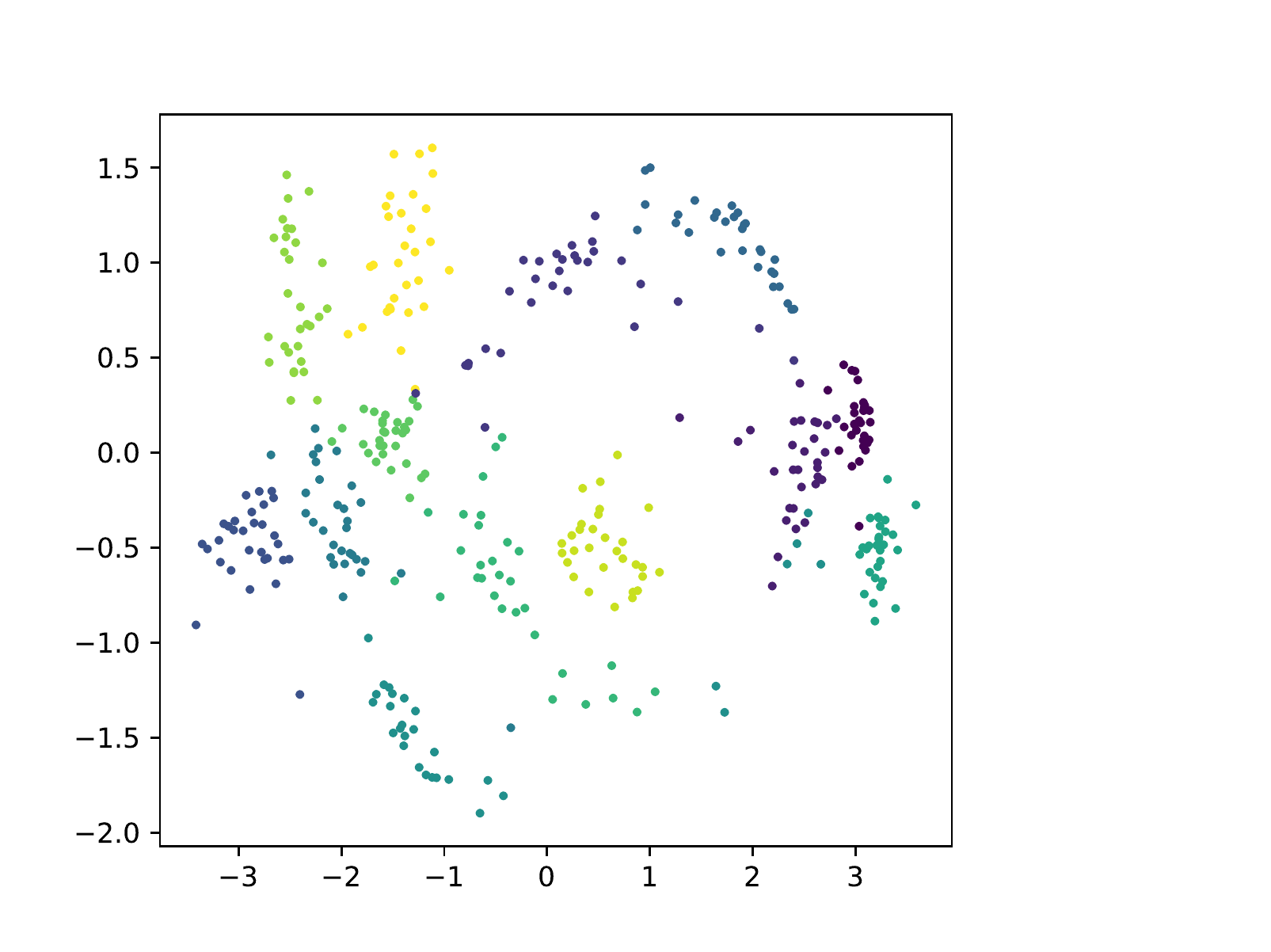}}
\end{minipage}
\centering
\begin{minipage}{0.27\linewidth}
    \subfigure[GST-MLTTS]{\includegraphics[scale=0.35]{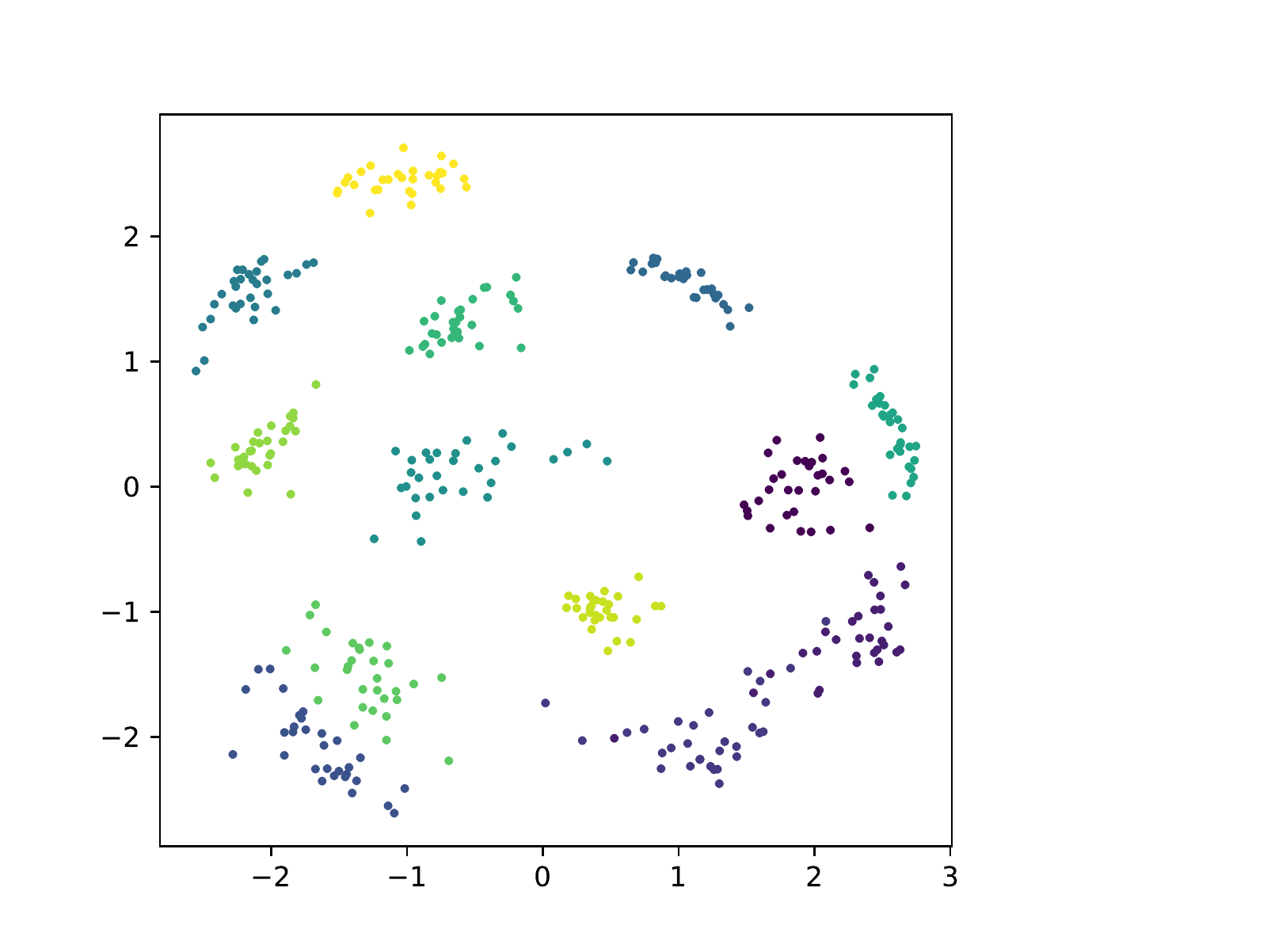}}
\end{minipage}
\centering
\begin{minipage}{0.27\linewidth}
    \subfigure[PromptStyle (Proposed)]{\includegraphics[scale=0.35]{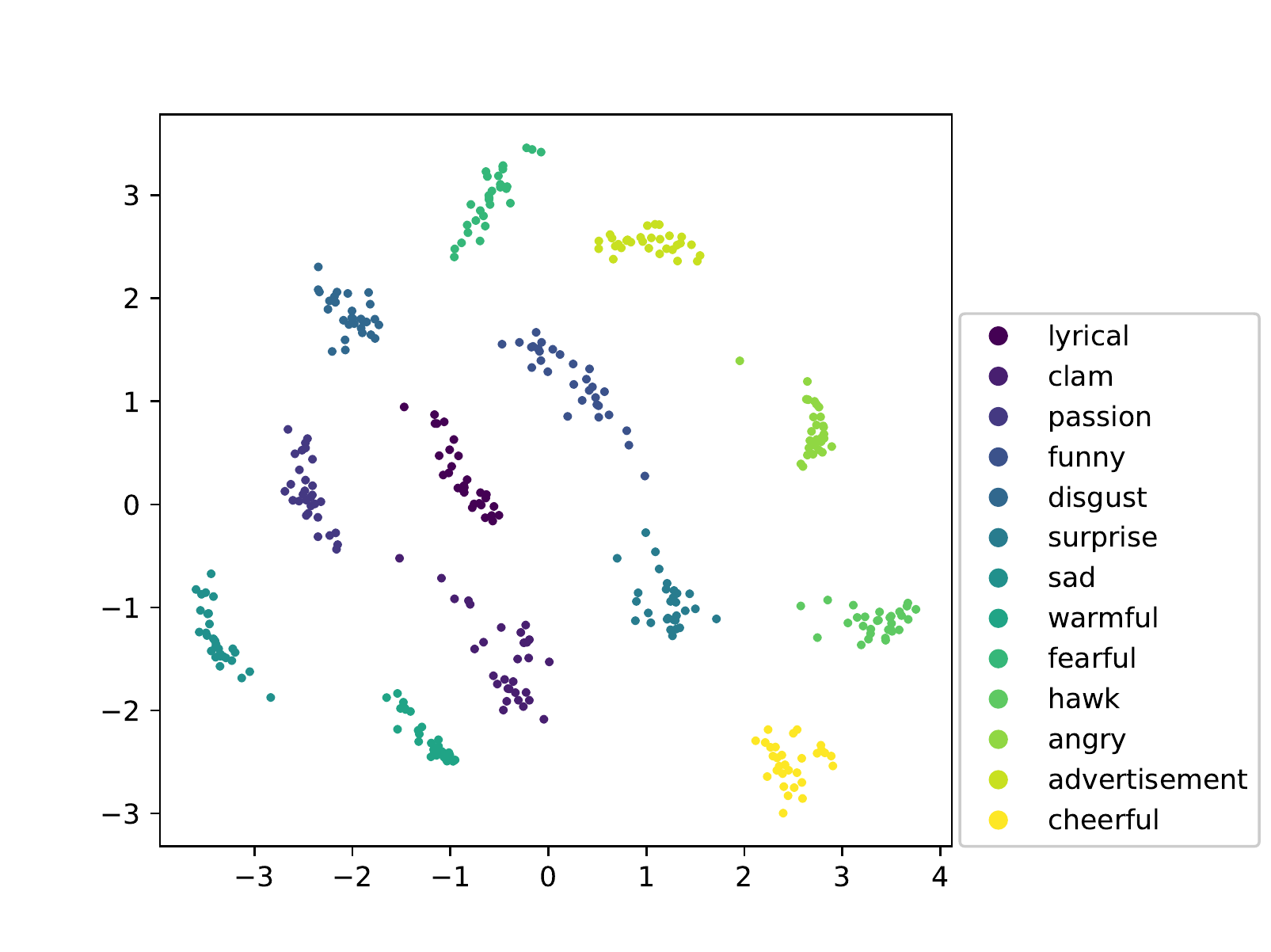}}
\end{minipage}\\
\caption{Visualization of the style embeddings from different models -- (a) CST-TTS, (b) GST-MLTTS and (c) PromptStyle.}
 \label{fig:tsne}
 \vspace{-0.4cm}
\end{figure*}

\vspace{-0.2cm}
\subsection{Training and Inference}
\vspace{-0.1cm}
We introduce a two-stage training procedure to obtain a style transfer model with text prompts. This bi-modal training approach has the following two advantages. 1) By establishing the relationship between the style and prompt spaces, either reference speech or text descriptions can be used to control the style transfer. 2) The type and scale of the annotated data can be determined specifically according to the actual domain requirements, and the fine-tuning process is time-saving. 

\vspace{-0.1cm}
\subsubsection{Stage 1: Style transfer by reference Mel}
\vspace{-0.05cm}

In the first stage, a style transfer system with reference speech is developed. We particularly study the role of each main module in VITS for the style transfer task. The text encoder processes the input phonemes, independent of the speaker and style information. The normalizing flow improves the flexibility of the prior distribution. The stochastic duration predictor estimates the distribution of phoneme duration from text embedding. We believe that duration is an important aspect of style, so only style embedding is added to the input of the stochastic duration predictor. For simplicity, the normalizing flow is designed to be a volume-preserving transformation with the Jacobian determinant of one. So we only add style embedding to the normalizing flow through global conditioning, considering that it is hard to take into account both speaker and style embedding in the construction of the prior distribution. The posterior encoder produces the normal posterior distribution from a linear spectrogram and the decoder reconstructs the latent variables $z$ to spectrogram. Referring to DelightfulTTS~\cite{liu2021delightfultts}, SSIM loss $L_{ssim}$ is added, and the speaker classifier in the adversarial component is trained with a cross-entropy loss with gradient inversion $L_{adv}$. Specifically, the loss of the first stage is 

\vspace{-0.1cm}
\begin{equation}
\begin{aligned}
L_{stage1} &= L_{vits}  + L_{ssim} + L_{adv}\label{1}. \\
\end{aligned}
\end{equation}
\vspace{-0.5cm}

\subsubsection{Stage 2: Style transfer by text prompt}
\vspace{-0.05cm}

The second stage aims to achieve controllable style transfer with text prompts. To this end, we use a prompt encoder to extract prompt embeddings from natural language descriptions, and the prompt embedding learns style embedding by cosine similarity loss referring to $L_{cons}$. The cosine similarity loss is not only used by a separate optimizer for the prompt encoder but also used to train the TTS system. Note that, we only train the last attention layer of BERT, and the parameters of the style encoder are frozen throughout this training process. The loss of the second stage is

\vspace{-0.1cm}
\begin{equation}
\begin{aligned}
L_{stage2} &= L_{vits}  + L_{cons}\label{2}. \\
\end{aligned}
\end{equation}
\vspace{-0.4cm}

\section{Experiments}

\subsection{Datasets}
\label{sec:datasets}
\vspace{-0.1cm}

We use 2 hours of an audiobook corpus recorded from a single speaker as the target speaker. Most recordings are common reading style without much expressive speech. As there is no public expressive dataset with abundant natural language prompts, we use an internally expressive Mandarin Chinese speech corpus as source speakers. The corpus contains 12 hours of speech data from 8 female speakers and about 6 hours of speech is associated with natural language descriptions. To obtain the natural language descriptions (text prompts), we 
invite 9 professional annotators to use a phrase or a sentence to describe the style of each utterance. They are told to focus on the speaking style and ignore the linguistic content.

For our experiments, all the speech recordings are downsampled to 24 kHz. Linear spectrograms obtained from raw waveforms through the Short-time Fourier transform (STFT) are used as the input of VITS. Eighty-dimensional Mel-frequency spectrograms are extracted with 12.5 ms frame shift and 50 ms frame length. The text prompts are tokenized by a pre-trained BERT.

\vspace{-0.2cm}
\subsection{Model Configuration}
\vspace{-0.1cm}

Our work focuses on style transfer with text prompts based on the audiobook datasets. Thus two models of style transfer on audiobook datasets are established as baselines -- one with text prompt and the other using reference speech.

\begin{itemize}
    \item \textbf{CST-TTS}: Following~\cite{shin2022text}, a cross-speaker style transfer framework is built based on non-autoregressive feedforward structure which can control the expressiveness by style tags.

\item \textbf{GST-MLTTS}: A cross-speaker emotion transfer model~\cite{Wu2019EndtoEndES} based on semi-supervised training and SCLN.

\item \textbf{PromptStyle}: The proposed style transfer system. The hyperparameters of VITS are set to be the same as in \cite{kim2021conditional}. The reference encoder has the same architecture as GSTs without style tokens. The prompt encoder uses a pre-trained BERT to extract 768-dimensional global semantic embedding, and an adapter layer consisting of three linear layers with ReLU activation converting the semantic embedding to 256-dimensional prompt embedding.
\end{itemize}

All models are trained with a batch size of 32 for 200k steps, while the fine-tuning stage of \textit{StylePrompt} is trained with 16 batches for 10k steps. HiFi-GAN~\cite{kong2020hifi} is used in \textit{CST-TTS} and \textit{GST-MLTTS} as the neural vocoder for reconstructing the waveform.

\vspace{-0.1cm}
\subsection{Performance on the Style Transfer}
\subsubsection{Subjective Evaluation}
The performances of three comparison models on style transfer are evaluated in terms of speech quality, speaker similarity, and style similarity with a Mean Opinion Score (MOS) test. Specifically, 50 speech samples are synthesized for each model, and 15 listeners who are native Chinese take part in this experiment. They are asked to score on a 1$\sim$5 point with 0.5 intervals for the speech quality, speaker similarity with the target speaker, and style similarity with the source speaker. As shown in Table~\ref{table1:sub-eval-trans}, PromptStyle has clearly better performance in terms of speech quality, speaker similarity, and style similarity. We believe such good performance is due to the end-to-end structure and the style embedding as well as the speaker embedding added to different modules as conditions.

\begin{table}[t]
\centering
\caption{Similarity/Quality MOS results (Average score and 95\% confidence interval)}

\begin{tabular}{cccc}
\hline
Models         & \begin{tabular}[c]{@{}c@{}}Speakers\\ Similarity\end{tabular} & \begin{tabular}[c]{@{}c@{}}Style\\ Similarity\end{tabular}  & \begin{tabular}[c]{@{}c@{}}Speech\\ Quality\end{tabular}   \\ \hline
CST-TTS        &      3.01      &       3.23      &     3.14                \\
GST-MLTTS      &      3.46      &       3.42      &     3.53                \\
PromptStyle    &      \textbf{3.63}      &       \textbf{3.78}      &     \textbf{4.06}                \\ \hline
\end{tabular}
\vspace{-0.05cm}
\label{table1:sub-eval-trans}
\end{table}

\begin{table}[t]
 \caption{The setup of the ablation experiments. }
\label{tab1:data}
\centering
\setlength{\tabcolsep}{1mm}
\begin{tabular}{@{}c|cccccc@{}} \hline
\multirow{2}{*}{Models} &\multicolumn{2}{c}{Posterior Encoder}  &\multicolumn{2}{c}{Flow} &\multicolumn{2}{c}{Decoder}  \\ \cline{2-7}
& style & speaker & style & speaker & style & speaker  \\ 
\midrule
M1 & \checkmark & \checkmark & \checkmark & \checkmark & \checkmark & \checkmark\\
\midrule
M2 & \checkmark & \checkmark & \checkmark &  & \checkmark & \checkmark\\
\midrule
PromptStyle & \checkmark & \checkmark & \checkmark &  &  & \checkmark\\
\midrule
M3 &  & \checkmark & \checkmark &  &  & \checkmark\\\hline
\end{tabular}
\vspace{-0.4cm}
\label{ablation_setup}
\end{table}

\subsubsection{Objective evaluation}
%we manually select 30 utterances for the 13 styles in the expressive database without style annotations.
To visualize style embedding space by t-SNE~\cite{Laurens2008Visualizing}, We manually selected 13 common styles from the training dataset for evaluation. Style embedding is extracted from the selected speech using PromptStyle, ST-TTS, and GST-MLTTS respectively. As shown in Fig~\ref{fig:tsne}, each point indicates a style embedding and points with the same color are from the same style category. The distance between the two points indicates the relative similarity of the embeddings. Smaller distances indicate more similar embeddings. An ideal style encoder should make the style embeddings from the same category closer to each other while embeddings from different categories should be separated apart. From Fig~\ref{fig:tsne}, we observe that MST-TTS gets the worst performance and GST-MLTTS performs better but the points from the same category are still diversely distributed. In contrast, in Fig.~\ref{fig:tsne}(c), the style embeddings from the same category are clustered together, and different clusters are apart from each other. It indicates that PromptStyle presents significant superiority in extracting style embeddings.

\subsection{Ablation Study}

In order to verify the influence of the style embedding and speaker embedding added to different modules as conditions, we conduct the ablation experiments considering the posterior encoder, the normalizing flow, and the decoder. Intuitively, we add both style embeddings and speaker embeddings to all their modules, referring to \textbf{M1}. Then we remove speaker embeddings from the flow, referring to \textbf{M2}. On the base of M2, removing style embeddings from the decoder is our final version of StylePrompt. Finally, on the base of StylePrompt, we remove the style embeddings from the posterior encoder, referring to \textbf{M3}. Similarly, these models are evaluated in terms of speech quality, speaker similarity, and style similarity with a MOS test. The model setups and experimental results are shown in Table~\ref{ablation_setup} and Table~\ref{ablation_result}, respectively.

As shown in Table~\ref{ablation_result}, M1 maintains well the timbre of the target speaker but fails to achieve decent style transfer. Comparing M2 and M1, the results indicate that it is hard for a normalizing flow to extract a proper style representation with both style embedding and speaker embedding as conditions. Comparing PromptStyle with M2, it is verified that PromptStyle can capture more speaker timbre when the speaker embedding is added to the decoder of VITS. Finally, comparing M3 with PromptStyle, we find that adding the style embedding to the posterior encoder as conditions is beneficial to improve the expressiveness of the synthesized speech.

\begin{table}[t]
\centering
\caption{Similarity/Quality MOS results of ablations (Average score and 95\% confidence interval)}

\begin{tabular}{cccc}
\hline
Models         & \begin{tabular}[c]{@{}c@{}}Speakers\\ Similarity\end{tabular} & \begin{tabular}[c]{@{}c@{}}Style\\ Similarity\end{tabular}  & \begin{tabular}[c]{@{}c@{}}Speech\\ Quality\end{tabular}   \\ \hline
M1           &        \textbf{3.96}            &     2.87        &       4.07              \\
M2           &        3.43            &     3.78        &       3.98              \\
PromptStyle           &        3.62        &     \textbf{3.82}        &      4.05               \\
M3     &      3.57              &     3.73        &      4.01               \\ \hline
\end{tabular}
\label{ablation_result}
\end{table}

\begin{figure}[t]
	\centering
	\includegraphics[scale=0.45]{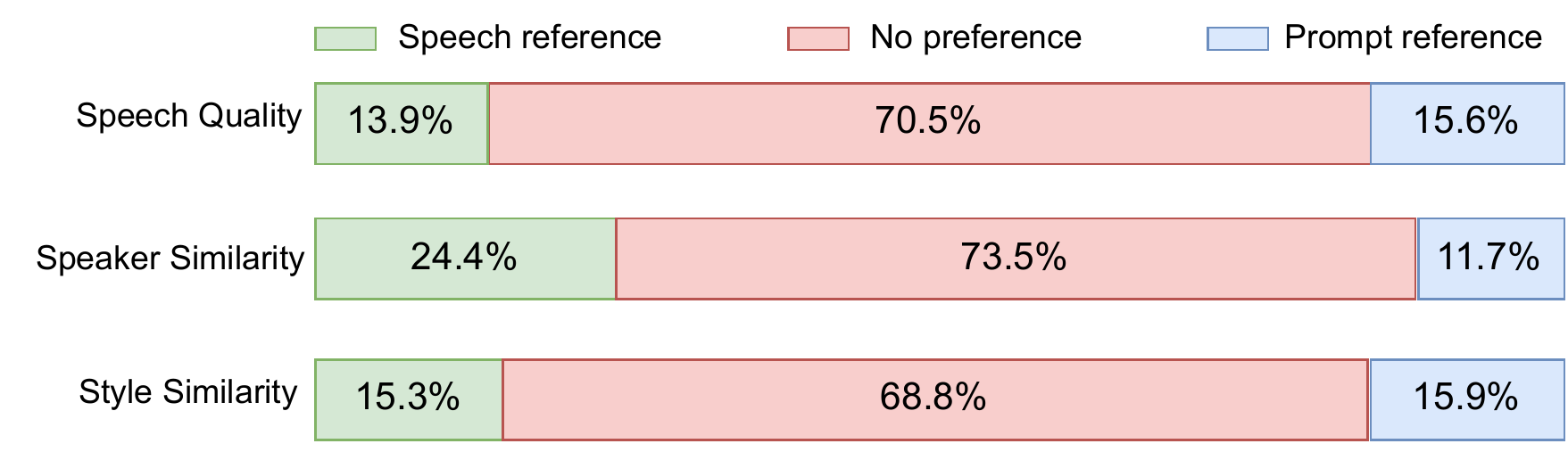}
	\caption{ABX preference results}
	\vspace{-0.4cm}
 \label{fig:abx}
\end{figure}

\subsection{Style Control by Text Prompt}
%We conduct ABX preference tests on speech quality, speaker similarity, and style similarity to evaluate the ability of PromptStyle on controlling transferred style with text prompts. 

To evaluate the performance of PromptStyle in style transfer by text prompts, we further conduct ABX preference tests between the prompt embedding and the style embedding in terms of speech quality, speaker similarity, and style similarity. Note that style embedding serves as the performance top line, as speech reference contains more direct style information than the text prompt. Precisely, we extract style embedding and prompt embedding from the reference speech and the text prompt, respectively, to control the transferred style. Listeners are offered a pair of randomly selected samples and asked to choose which one is as expressive as the descriptions of the provided text prompts. As the preference test results shown in Fig.~\ref{fig:abx}, the listeners give higher preference to \textit{No preference}, which means that our proposed method achieves the goal of text prompt guided style transfer with precise style control and high speech quality.

\section{Conclusions}
In this work, we propose PromptStyle, a style transfer model with natural language prompts, to make the style transfer more controllable and user-friendly. PromptStyle establishes the relationship between the style embedding space and the prompt embedding space through two-stage training. Experiments on expressive audiobook synthesis show that PromptStyle can achieve the goal of style transfer with text prompts while maintaining relatively high stability and speaker similarity. %In the future, we will explore to imporve the generalization capability of PromptStyle to extract better representations from prompts.

% \subsubsection{Subjective Evaluation}

% \subsubsection{Objective Evaluation}

\clearpage

\bibliographystyle{IEEEtran}
\bibliography{mybib}

\end{document}